\documentclass[12pt]{article}

\usepackage{amsfonts}
\usepackage{amssymb}

\newcommand{\beq}[1]{\begin{equation}\label{#1}}
\newcommand{\eeq}{\end{equation}}
\newcommand{\bear}[1]{\begin{eqnarray}\label{#1}}
\newcommand{\ear}{\end{eqnarray}}

\catcode`\@=11 \@addtoreset{equation}{section}\catcode`\@=12

\newcommand{\np}{ {\newpage } }

 \newcommand{\R}{ {\mathbb R} }

\newcommand{\eps}{ \varepsilon }

 \newcommand{\fnm}{\footnotemark}
 \newcommand{\fnt}{\footnotetext}

\begin{document}

\begin{center}    \bf \large

On supersymmetric  M-brane   configurations with  $\R^{1,1}_{*}/Z_2$ submanifold

\end{center}

\vspace{0.96truecm}

\begin{center}

 \normalsize\bf V. D. Ivashchuk\fnm[1]\fnt[1]{e-mail:  ivashchuk@mail.ru}

\vspace{0.3truecm}

 \it Center for Gravitation and Fundamental Metrology,
 VNIIMS, 46 Ozyornaya ul., Moscow 119361, Russia, \\

 \it Institute of Gravitation and Cosmology,
 Peoples' Friendship University of Russia,
 6 Miklukho-Maklaya ul., Moscow 117198, Russia

\end{center}

\begin{abstract}

We obtain new examples of partially supersymmetric $M$-brane  solutions defined on  products of Ricci-flat
manifolds, which contain two-dimensional Lorentzian submanifold $\R^{1,1}_{*}/Z_2$ with one parallel spinor.
The examples belong to the following configurations:  $M2$, $M5$, $M2 \cap M5$ and $M5 \cap M5$.
Among them a $M2$ solution  with ${\cal N} =1/32$ fractional number of preserved supersymmetries  is presented.
The examples with three $M$-branes were considered earlier in our paper with A.A. Golubtsova.

\end{abstract}

\np

\section{\bf Introduction}
\setcounter{equation}{0}

In this letter  we present new examples  of (partially) supersymmetric solutions
in 11-dimensional supergravity \cite{CJS}.  The renewed interest
to supersymmetric solutions may be motivated by the developments  of \cite{BL1,Gust,ABJM} (for recent
publications see  \cite{BJLRZ} and refs. therein)
 and also by proving new supergravity/gauge (or AdS/CFT) correspondences: for a review see \cite{Hub} and refs. therein, e.g.  pioneering papers:  \cite{Mal} and   \cite{GKP,Witten}.

We deal with the following $M$-brane  configurations: $M2$, $M5$, $M2 \cap M5$ and $M5 \cap M5$
defined on products of (spin) Ricci-flat manifolds  and use our general approach to $M$-brane solutions
(e.g. supersymmetric ones) from \cite{IM0,IM01,Iv-00,Iv-12,Gol-Iv-1,Gol-Iv-2}, which was inspired by earlier
(e.g. pioneering) papers \cite{DS}-\cite{FOF}.

The examples of  supersymmetric solutions,  presented here, are defined on eleven-dimensional
manifold which contain a  two-dimensional Lorentzian submanifold $\R^{1,1}_{*}/Z_2$
with one parallel spinor.

Here and in what follows  $\R^{1,1}$ is the  2-dimensional pseudo-Euclidean space
 $(\R^{2},\eta)$ with the metric $\eta = - d y^0  \otimes d y^0 + d y^1 \otimes d y^1$.
By $\R^{1,1}_{*}/Z_2 = (\R^{1,1}_{*}/Z_2)_{\eps}$  we denote a spin manifold
 \cite{Baum},   which has one of two inequivalent ($Spin(1,1)$) spin structures labeled
by index $\eps = +1, -1$. We use (e.g. as a base of the spin manifold) the  quotient space
$\R^{1,1}_{*} = \R^{1,1} \setminus \{(0,0)\}$ (which as a set is coinciding with
$\R^2_{*} = \R^2 \setminus \{(0,0)\}$) with the metric induced from
$\R^{1,1}$.  The generator of $Z_{2}$ acts on $\R^{1,1}_{*}$ (or $\R^2_{*}$) by $(x_1,x_2) \mapsto
   (-x_1,-x_2)$.  The spin manifold $ (\R^{1,1}_{*}/Z_2)_{\eps}$ has one parallel (i.e. covariantly constant) chiral spinor,
   or more rigorously, the pair of dimensions of spaces of chiral parallel spinors   $(n(1), n(-1))$  is either
$(1,0)$ (for $\eps = +1$), or $(0,1)$ (for $\eps = -1$).

It should be noted that  supersymmetric configurations with
three $M$-branes, were considered earlier in \cite{Gol-Iv-1,Gol-Iv-2};
some examples which contain  $\R^{1,1}_{*}/Z_2$ submanifold were presented
in \cite{Gol-Iv-2}. 

In this paper we use notations from \cite{Iv-12,Gol-Iv-2}. For any
(spin) manifold $M_i$ ($i=0,...,n$, where $n$ is the number of
factor spaces) we denote by $n_i(c_{(i)})$ the number of (linear independent) chiral
parallel spinors $\eta_i$ with chiral number $c_{(i)}$, satisfying
$\hat{\Gamma}_{(i)} \eta_i  = c_{(i)} \eta_i$, where
$\hat{\Gamma}_{(i)}$ is the product of (frame) gamma-matrices
corresponding to $M_i$ and by $n_i$ - the number of parallel
spinors on $M_i$. For oriented (e.g. spin) manifold $M_i$ with the metric
$g^i$ we denote by $\tau_i = \tau_i(g^i)$ the volume form corresponding to $g^i$.

\section{\bf Solutions with one brane}

 \subsection{$M2$-brane}

Let us we consider  the electric  $2$-brane solution
defined on the manifold

 \beq{3.2a}
 M_{0}  \times M_{1}.
 \eeq

The solution reads \cite{Iv-00,Iv-12}

 \bear{3.2}
  g= H^{1/3} \{ g^0  + H^{-1} g^1   \},
  \\ \label{3.3}
  F = c d H^{-1} \wedge \tau_1,
 \ear
 where  $c^2 = 1$, $H = H(x)$ is a harmonic function on
 $(M_0,g^0)$,   $d_0 = 8$, $d_1 = 3$ and the metrics $g^i$, $i =
  0,1$, are Ricci-flat; $g^0$ has the Euclidean signature and $g^1$ has
 the signature $(-,+,+)$.

The number of unbroken SUSY is at least \cite{Iv-00}

 \beq{3.10} {\cal N} = n_0(c) n_1/32, \eeq
where  $n_0(c)$ is the number of chiral parallel  spinors on $M_0$
and $n_1$ is the number of parallel spinors on $M_1$.

Since  $(M_1,g^1)$ is  $3$-dimensional Ricci-flat space, it is flat,
i.e. the Riemann tensor corresponding to $g^1$ is zero.

Let us put here  $M_1 = (\R^{1,1}_{*}/Z_2) \times \R $. We get $n_1 = 1$ and hence
 \beq{3.11a}
 {\cal N} = n_0(c)/32.
 \eeq

 {\bf Example 1.}
According to M. Wang's classification \cite{Wang}  an irreducible,
simply-connected,  Riemannian $8$-dimensional   manifold $M_0$
admitting parallel spinors must have precisely one of the
following holonomy groups $H$: $Spin(7)$, $SU(4)$ or $Sp(2)$ .
 For suitably chosen  orientation on $M_0$ the numbers of chiral
parallel spinors are the following ones: $(n_0(+1), n_0(-1)) = (k,
0)$, where $k = 1, 2, 3$ for $H  = Spin(7), SU(4),  Sp(2)$,
respectively. Hence  ${\cal N} = k /32$ for $c = + 1$ and ${\cal
N}  = 0$ for $c = -1$.

 Thus, for $c = + 1$ and $k = 1$ we obtain an example of (partially) supersymmetric solution with  ${\cal N} = 1/32$
fractional number of unbroken SUSY.
The solution  is defined on the product of Riemannian $8$-dimensional  manifold $M_0$ of holonomy $Spin(7)$ and
 flat $3$-dimensional manifold $M_1 = (\R^{1,1}_{*}/Z_2) \times \R $ of pseudo-Euclidean  signature $(-,+,+)$.

 {\bf Example 2. }
Let us consider $K3 = CY_2$ which is a 4-dimensional  Ricci-flat
K\"{a}hler manifold with the holonomy group $SU(2) = Sp(1)$ and
self-dual (or anti-self-dual) curvature tensor. $K3$ has two
Killing spinors of the same chirality, say, $+1$.   Let $M_0 =
\R^4 \times K3$, then we get $n_0 (+1) = n_0 (-1) = 4$ and hence
 ${\cal N} =  1/8$ for any $c = \pm 1$.
  For  $M_0 = K3 \times K3$,  we obtain $n_0 (+1) = 4$,  $n_0 (-1) = 0$
 and hence ${\cal N} =  1/8$ for  $c = 1$ and ${\cal N} = 0$  for  $c = -1$ .

{\bf Example 3. }
 Let $M_0 = \R^2 \times CY_3$, where  $CY_3$ is 6-dimensional Calabi-Yau manifold (3-fold) of
 holonomy $SU(3)$. Since $CY_3$ has two parallel spinors of
 opposite chiralities (and  the same for $\R^2$)  we  get $n_0 (+1) = n_0 (-1) = 2$
  and hence ${\cal N} =  1/16$ for any $c = \pm 1$.

\subsection{$M5$-brane}

Now we consider the magnetic  $5$-brane solution defined on the
manifold  (\ref{3.2a}) with $d_0 =5$ and $d_1 = 6$ \cite{Iv-00,Iv-12}:

 \bear{3.4m} g= H^{2/3} \{ g^0  + H^{-1} g^1 \},
 \\ \label{3.5m}
 F = c (*_0 d H),
 \ear
 where $c^2 = 1$,
 $*_0$ is the Hodge operator on $(M_0,g^0)$,
 $H = H(x)$ is a harmonic function on
$(M_0,g^0)$,  and metrics $g^i$, $i = 0,1$, are Ricci-flat, $g^0$
has the Euclidean signature and $g^1$ has the signature
$(-,+,+,+,+,+)$.

 The number of unbroken SUSY is at least \cite{Iv-00}

 \beq{3.10m} {\cal N} = n_0 n_1(c)/32, \eeq
where  $n_0$ is the number of  parallel spinors on $M_0$  and
$n_1(c)$ is the number of chiral parallel spinors on $M_1$ with the chirality $c$.

{\bf Example 4.} Let $M_1 = (\R^{1,1}_{*}/Z_2) \times
 K3$. For a suitable choice of orientations  $n_1(1) = 2$ and  $n_1(-1) = 0$.
 In this case  we obtain from (\ref{3.10m}):  ${\cal N} = n_0/16$    for $c = 1$ and
  ${\cal N} = 0$ for $c = -1$.
 For flat $M_0 = \R^5$ (with $n_0 =
 4$) we get   ${\cal N} = 1/4$ for $c = 1$ . For $M_0 = \R \times K3$
 we have $n_0 = 2$ and hence ${\cal N} = 1/8$ for $c = 1$.

{\bf Remark.} Let $n(M)$ is the number of parallel spinors on the (spin) manifold 
 $M$and  $n(c,M)$ is the number of chiral parallel spinors on the (spin) even-dimensional manifold 
 $M$ with the chiral number $c$. In this paper we use implicitly the following two relations:
 i) $n(M_0) = n(\R \times M_0)$; ii)  $n(c, M_1 \times M_2) = \sum_{c_1 c_2 = c} n(c_1,M_1) n(c_2, M_2)$;
  where $M_0, M_1, M_2$ are even-dimensional manifolds. These relations may be readily proved just
 by using the formalism of \cite{Iv-12}. Relation ii) implies:  $n(M_1 \times M_2) = 
  n(M_1) n(M_2)$.

\section{\bf Solutions with two branes}

\subsection{$M2 \cap M5$-branes}

Let us consider a solution  with intersecting $M2$- and
$M5$-branes defined on the manifold

 \bear{4.1}
 M_{0}  \times M_{1}  \times M_{2} \times M_{3},
  \ear
 where $d_0 = 4$, $d_1 = 1$, $d_2 = 4$ and $d_3 = 2$.

 The solution reads \cite{Iv-12}

 \bear{4.2}
 g= H_1^{1/3} H_2^{2/3}
 \{ g^0  + H_1^{-1} g^1 +  H_2^{-1} g^2
 + H_1^{-1} H_2^{-1} g^3 \},
 \\ \label{4.3}
 F = c_1 d H^{-1}_1 \wedge  \tau_1 \wedge \tau_3 +
     c_2 (*_0 d H_2) \wedge \tau_1,
 \ear
where $c^2_1 = c^2_2 = 1$; $H_1, H_2$ are harmonic functions on
$(M_0,g^0)$,   metrics $g^0, g^2$ are Ricci-flat and $g^1, g^3$
are flat. The metrics $g^i$, $i = 0,1,2$, have Euclidean signatures
and the metric $g^3$ has the signature $(-,+)$.  We put here $M_1 = \R$.

  The number of unbroken SUSY is  given by \cite{Iv-12}

  \beq{4.9a}
    {\cal N} = \frac{1}{32} ( n_0( c_1)  n_2(1) n_3(c_2)
          +  n_0(- c_1) n_2(-1)  n_3(- c_2) ),
   \eeq
 where $n_j (c_{(j)})$ is the number of chiral parallel spinors on $M_j$ with
 the chirality number $c_{(j)}$, $j = 0,2,3$.

  We put  $M_3 = \R^{1,1}_{*}/Z_2$ and $(n_3(1), n_3(-1)) = (1, 0)$.   Then  we get from (\ref{4.9a})

    \beq{4.11a}
      {\cal N} =  \frac{1}{32}  n_0( c_1 c_2 )  n_2( c_2).
     \eeq
            
      {\bf Example 5.}
      For flat $M_0 = M_2 = \R^4$ we get
      ${\cal N} = 1/8$ for any $c_2$.

   {\bf Example 6: one $K3$ factor-space.} Let $M_0 = \R^4$ and $M_2 = K3$ with
         $(n_{2}(1), n_{2}(-1)) =  (2,0)$. We get
   from (\ref{4.11a}): ${ \cal N } = 1/8$ for $c_2 = + 1$ and ${\cal N} = 0$
         for $c_2 = -1$ ($c_1$ is arbitrary). 
   For  $M_0 = K3$ with $(n_{2}(1), n_{2}(-1)) =  (2,0)$ and $M_2 = \R^4$ 
   we obtain  from (\ref{4.11a}): ${ \cal N } = 1/8$ for $c_1 c_2 = + 1$ and ${\cal N} = 0$
            for $c_1 c_2 = -1$. 

   {\bf Example 7: two $K3$ factor-spaces.} Let $M_0 =  M_2 = K3$.
    We put for chiral numbers   $(n_i(+1), n_i(-1)) = (2, 0)$, $i =
    0,2$. Then we get ${\cal N} = 1/8$ for $c_1 = c_2 = + 1$ and ${\cal N} = 0$
    otherwise.

   \subsection{$M5 \cap M5$-branes}

Now we deal with $M5 \cap M5$-solution defined on the manifold
(\ref{4.1})  with $d_0 = 3 $, $d_1 = d_2 = 2$ and $d_3 = 4$.

 The solution reads \cite{Iv-12}
 \bear{6.2}
 g= H_1^{2/3} H_2^{2/3}
 \{ g^0  + H_1^{-1} g^1 +  H_2^{-1} g^2  + H_1^{-1} H_2^{-1} g^3 \},
 \\ \label{6.3}
 F = c_1   (*_0 d H_1) \wedge \tau_2 +
     c_2   (*_0 d H_2) \wedge \tau_1,
 \ear
where $c^2_1 = c^2_2 = 1$; $H_1$, $H_2$ are harmonic functions on
$(M_0,g^0)$,   metrics $g^i$, $i = 0,1,2,3$, are Ricci-flat ( the
first three metrics are flat). The metrics $g^i$, $i = 0,1,2$,
have Euclidean signatures and the metric $g^3$ has the signature
 $(-,+,+,+)$.

The fractional number of preserved SUSY is given by \cite{Iv-12}
  \beq{6.9a}
   {\cal N} = \frac{1}{32} (n_0  n_1(-i c_1)  n_2(-i c_2) n_3(+i)
             +  n_0  n_1( i c_1)  n_2( i c_2) n_3(-i)) .
   \eeq

 Here $n_j (c_{(j)})$ is the number of chiral parallel spinors on $M_j$,
 $j = 1,2,3$, and $n_0$ is the number of   parallel spinors on $M_0$.

  We put $M_1 = M_2 = \R^2$.
     We get
    $n_1 (c) = n_2 (c) = 1$ and hence

   \beq{6.11}
    { \cal N} = \frac{1}{32} n_0  n_3.
  \eeq

  {\bf Example 8. } 
  For flat factor-space $M_0 = \R^3$  and
  $M_3 = (\R^{1,1}_{*}/Z_2) \times \R^2$  we get $n_0 = n_3 = 2$. 
  Thus, we are led to the fractional number  ${\cal N} = 1/8$. It is twice less than
  the number ${\cal N} = 1/4$ corresponding to the  case $M_3 = \R^{1,3}$ \cite{BREJS}.

\section{Conclusions and discussions}

In this letter we have obtained new examples of (partially) supersymmetric
$M$-brane solutions  defined on the products of Ricci-flat (e.g. flat) manifolds,
which contain  two-dimensional  manifold $\R^{1,1}_{*}/Z_2$ with one parallel spinor.

These  examples contain  configurations with one and two branes:
$M2$, $M5$, $M2 \cap M5$ and $M5 \cap M5$.  Here we have found an example
of $M2$-brane  solution with the minimal number of  fractional supersymmetries equal to  $1/32$.
This solution is defined on the product of  8-dimensional Riemannian manifold  of $Spin(7)$-holonomy
admitting one parallel spinor and 3-dimensional
flat Lorentzian manifold $M_1 = (\R^{1,1}_{*}/Z_2) \times \R$ also admitting one parallel spinor.

Here we have considered the solutions with generic harmonic functions.
One topic of interest may be in analyzing of special
``near-horizon'' solutions with moduli harmonic  functions  defined
on  cones over certain Einstein spaces. In the
``near-horizon'' case the fractional numbers of unbroken
supersymmetries might be larger (e.g. twice larger) then ``at
least'' numbers ${\cal N}$ obtained here for generic
harmonic functions. Such solutions which are a sort of  Freund-Rubin-type solutions with
composite $M$-branes (see \cite{I-2} and references therein),  may of interest in a context
of AdS/CFT approach and its modifications.
Another  topic of interest may be in analyzing of  partially supersymmetric solutions for $IIA$, $IIB$
and other supergravity models in various dimensions, which contain two-dimensional
submanifold $\R^{1,1}_{*}/Z_2$.

The Lorentzian manifold $\R^{1,1}_{*}/Z_2$ with the golonomy group $H = Z_2$  is not geodesically complete.
It may be considered as a smooth interior part of $2d$ Lorentzian  orbifold  $\R^{1,1}/Z_2$.
An open problem here is to use this orbifold  in  supertring/M-theory applications. 
We note that earlier another Lorentzian  (null) orbifold - so-called ``Milne Universe'' - was used in a context of string theory, see \cite{HS,Nek} and other related publications.


\begin{center}
{\bf Acknowledgments}
\end{center}

The author is grateful to  D.V. Alekseevsky, H. Baum, D. Berenstein, L. Bonora, A. Bytsenko, A.A. Golubtsova, D.V. Gal'tsov, G. Papadopoulos  and K. Stelle    for useful discussions and/or informing on certain topics related to the subject of the paper. The author is also grateful to H. Nicolai for hospitality during the visiting of the Albert Einstein Institute in Golm (Jan.-Feb. 2015) where the decision of preparing this paper was adopted.

 \end{document}